\documentclass[useAMS]{mn2e}
\usepackage{times}
\input{epsf}

\title[Reprocessing in the outer disc of XTE J1817--330]
{Reprocessing of X-rays in the outer accretion disc of the black hole binary XTE J1817--330}

\author[M. Gierli{\'n}ski, C. Done and K. Page]
{Marek
Gierli{\'n}ski$^{1}\thanks{E-mail:Marek.Gierlinski@durham.ac.uk}$,
Chris Done$^{1}$
and Kim Page$^{2}$\\
$^1$Department of Physics, University of Durham, South Road,
Durham DH1 3LE, UK\\
}

\date{Submitted to MNRAS}
\pagerange{\pageref{firstpage}--\pageref{lastpage}} \pubyear{2005}

\begin{document}

\topmargin = -0.5cm

\maketitle

\label{firstpage}

\begin{abstract}

We build a simple model of the optical/UV emission from
irradiation of the outer disc by the inner disc and coronal
emission in black hole binaries. We apply this to the broadband
{\it Swift} data from the outburst of the black hole binary XTE
J1817--330 to confirm previous results that the optical/UV
emission in the soft state is consistent with a reprocessing a
constant fraction of the bolometric X-ray luminosity. However,
this is very surprising as the disc temperature drops by more
than a factor 3 in the soft state, which should produce a marked
change in the reprocessing efficiency. The easiest way to match
the observed constant reprocessed fraction is for the disc skin
to be highly ionized (as suggested 30 years ago by van Paradijs),
so that the bulk of the disc flux is reflected and only the
hardest X-rays heat the disc. The constant reprocessed fraction
also favours direct illumination of the disc over a scattering
origin as the optical depth/solid angle of any scattering
material (wind/corona) over the disc should decrease as the
source luminosity declines. By contrast, the reprocessed fraction
increases very significantly (by a factor $\sim$6) as the source
enters the hard state. This dramatic change is not evident from
X-ray/UV flux correlations as it is masked by bandpass effects.
However, it does not necessarily signal a change in emission e.g.
the emergence of the jet dominating the optical/UV flux as the
reflection albedo must change with the dramatic change in
spectral shape.

\end{abstract}


\section{Introduction} \label{sec:introduction}

Accreting black hole binaries radiate most of their energy in the
X-ray bandpass, and the wealth of recent X-ray data means that
their spectral behaviour in this energy range is well
characterized. At high accretion rates (typically more than a few
per cent of the Eddington luminosity) the accretion flow is
dominated by an optically thick and geometrically thin disc,
extending to the last stable orbit around the black hole. This
disc emits most of its power at $\sim$1 keV (Shakura \& Sunyaev
1973), though it is always accompanied by a soft tail of emission
to higher energies. By contrast, at lower luminosities the
spectrum is dominated by a hard tail, usually attributed to
Comptonization of soft seed photons in hot, optically thin
plasma, peaking at $\sim$100 keV. This soft-hard spectral
transition indicates a change in the nature and geometry of the
accretion flow, most plausibly due to the inner disc being
evaporated into a radiatively inefficient flow (see e.g. the
reviews by McClintock \& Remillard 2006; Done, Gierli{\'n}ski \&
Kubota 2007). Strong evidence for this picture comes from the
recent {\it Swift} monitoring campaign on the outburst of the low
mass X-ray binary XTE J1817--330 (Rykoff et al. 2007, hereafter
R07), where the low energy bandpass of the CCD detector gives for
the first time a direct detection of the disc during the soft to
hard transition and in the hard state (R07, Gierli{\'n}ski, Done
\& Page 2008, hereafter GDP08).  While the derived inner disc
radius is strongly model dependent in the hard state (simple fits
indicate that the disc is not truncated while more complex models
require a truncated disc: R07 and GDP08), all models give a
larger inner radius in the transition spectra than in the soft
state (R07, GDP08), as expected if a receding inner disc triggers
the soft-hard transition (GDP08).

Studies of the corresponding radio behaviour tie the jet firmly
into the accretion flow. The hard state has a steady jet whose
radio power increases with the X-ray luminosity of the flow. This
emission is suppressed in the soft, disc dominated state, but
there can be strong radio flaring during rapid transitions from
the hard to soft states. This is probably due to faster ejected
material (perhaps the remains of the jet supporting, large scale
height plasma from the low/hard state) catching up with the
slower, steady jet from lower luminosity states, giving rise to
strong shock acceleration (Fender, Belloni \& Gallo 2004; Done et
al. 2007).

By contrast with this wealth of data in the X-ray and radio
bandpass, the optical is surprisingly poorly studied. There is a
broad consensus that the reprocessing of the soft X-ray emission
from the outer disc should dominate the optical emission in the
soft state (e.g. van Paradijs \& McClintock 1994; Vrtilek et al.
1990; Esin et al. 2000; Hynes et al. 2002; 2005), but low/hard
state is much less clear. There can still be an important
contribution to the optical emission from reprocessed X-ray
illumination of the outer disc (e.g. Esin et al. 2000) but the
emission from the jet can also be important, as can the companion
star (e.g. Russell et al. 2006). Disentangling these in an
individual spectrum is not easy, but these models predict a
different optical to X--ray luminosity ratio, $L_{\rm O}/L_{\rm
X}$, from a series of spectra at different mass accretion rates.
Theoretical models of the optical emission from irradiation and
from the jet predict $L_{\rm O} \propto L_{\rm X}^{0.5}$ (van
Paradijs \& McClintock 1994) and $L_{\rm O} \propto L_{\rm
X}^{0.7}$ (Russell et al. 2006), respectively. However,
observations of a sample of low mass X-ray binaries (where the
companion star should make negligible contribution to the optical
emission except in quiescence) in the hard state shows $L_{\rm O}
\propto L_{\rm X}^{0.6}$ law, roughly consistent with either disc
reprocessing or the jet (Russell et al. 2006). Studies of the
correlated X-ray/optical rapid (subsecond) variability seem to
give a much better distinction between jet and disc models, since
the disc reprocessing model gives a clear prediction that the
optical light curve is a lagged and smeared version of the X-ray
light curve. The lack of smearing of the lagged optical
variability in the few cases where this has been attempted gives
clear evidence for a jet contribution (Malzac et al. 2004; Gandhi
et al. 2008).

Here we revisit the problem of disentangling the origin of the
UV/optical emission from snapshot spectra. We again use the {\it
Swift} data from the outburst of XTE J1817--330 (R07; GDP08) but
this time focus on the simultaneous optical/UV photometry which
accompanies each X-ray CCD spectrum. R07 showed that the UV flux
tracks the square root of the 2--10~keV X-ray flux in all the
data, and interpreted this as evidence for a reprocessing origin
in both the soft and hard states. We build a simple physical
model of reprocessing in the outer disc, and confirm the
conclusion of R07 that this produces the optical/UV flux.
However, we also show that simple flux-flux correlations mask a
clear increase in fraction of bolometric flux which is required
to be reprocessed and thermalized into the optical/UV as the
source enters the hard state. While this could signal a switch to
the jet dominating the optical/UV flux, we also show that this is
also consistent with the same constant illuminating geometry as
inferred for the soft state as the reflection albedo will change
markedly for the changing spectral shape. However, it does also
seem likely that there is a small jet contribution as well, since
this is required by weak cross-correlation signal between the
rapid optical and X-ray variability in a similarly bright hard
state of GX339-4 (Gandhi et al. 2008).

We caution that the flux correlations resulting from reprocessing
are not simple. More sophisticated physical models of X-ray
illuminated discs are required in order to clearly predict the
expected optical/UV flux from an irradiated disc, including the
self-consistent vertical structure of the illuminated disc
atmosphere, its reflection albedo and full emission spectrum
(e.g. Jiminez-Garate et al. 2002). Only with such models will
snapshot spectra be able to constrain the relative contribution
of jet and illuminated disc in the hard state, and be able
reliably estimate the solid angle of the directly illuminated
outer disc.

\section{Data reduction} \label{sec:data}

We use publicly available {\it Swift} data of the 2006 outburst
of XTE J1817--330. We use the same X-ray telescope (XRT) spectra
as in GDP08 (numbering follows that of R07, observation logs are
presented in their tables 1 and 2 for details). Additionally, we
extracted the corresponding UVOT data. The UVOT images were
aspect-corrected and {\sc uvot2pha} run to create {\sc
xspec}-compatible spectral points for every observation and
filter. In each case, a 5-arcsec circle was used to extract the
source counts, with a larger area used to estimate the
contribution from the background.

\begin{figure}
\begin{center}
\leavevmode  \epsfxsize=8.5cm \epsfbox{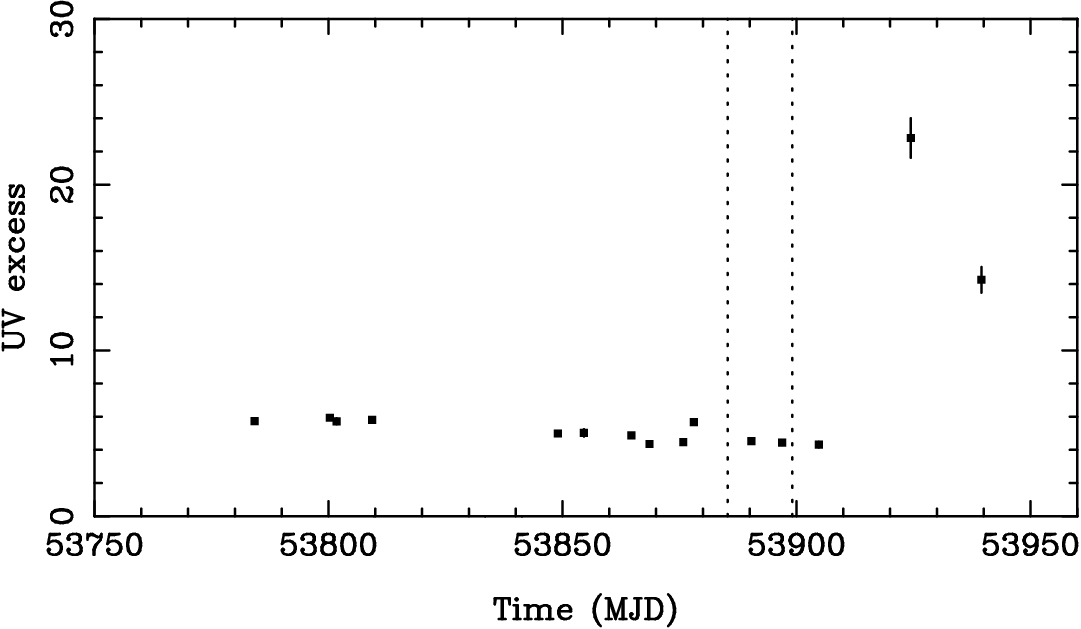}
\end{center}
\caption{UV excess: the ratio of the UVOT flux in UVW1 filter
divided by extrapolated {\sc diskbb} flux from best-fitting
simple disc models to XRT data. Dotted lines separate X-ray
spectral states: high/soft to the left, transition in the middle
and low/hard to the right (see GDP08).} \label{fig:excess}
\end{figure}

We first give an overview of the optical/UV data by extrapolating the
{\it Swift} XRT model fits to a standard disc ({\sc diskbb}) and
Comptonization ({\sc thcomp}) (detailed in section 3.2 and table 1 of
GDP08) down to the UVOT bandpass. We use the {\sc xspec} model {\sc
redden} to account for UV absorption, with fixed $E_{B-V}$ = 0.215 mag
(see R07). The X-ray fits always underpredict the optical/UV flux, and
we quantify this by the ratio of the observed to extrapolated flux in
the UVW1 filter of the UVOT.  This ratio is plotted as a function of
MJD in Fig. \ref{fig:excess}, showing that it remains relatively
constant throughout the outburst in the high/soft and transition
states (as defined in GDP08) but increases markedly in the low/hard
state. The excess is highly significant irrespective of uncertainties
in reddening e.g. in observation 1 the UV excess drops from 5.7 to 1.5
when $E_{B-V}$ is changed from 0.215 to 0 mag. In the next section we
build a physical model of the irradiated disc that can fit this
observed UV excess.

\section{Irradiated disc model} \label{sec:model}

\begin{figure}
\begin{center}
\leavevmode  \epsfxsize=8.5cm \epsfbox{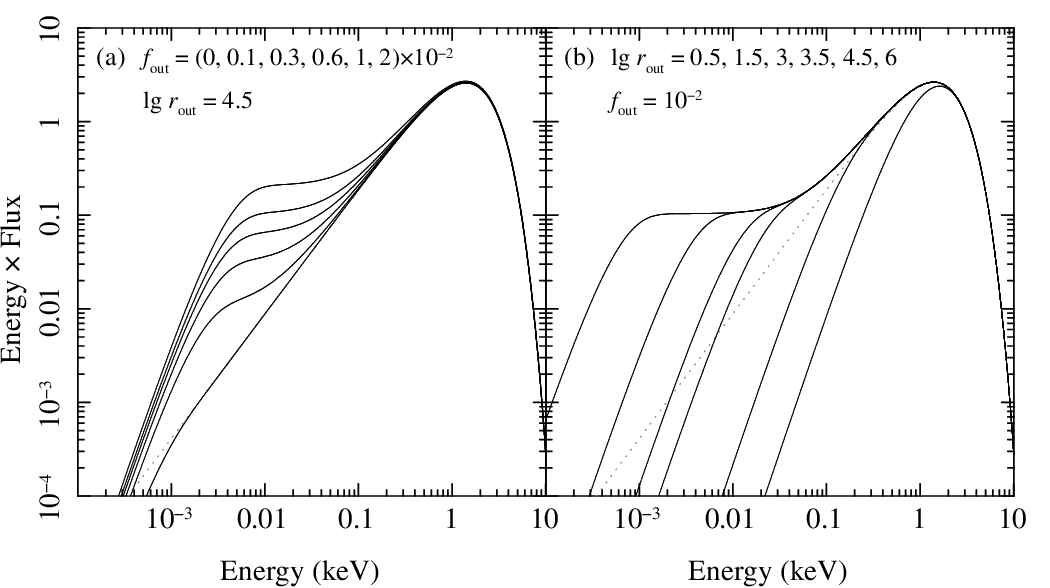}
\end{center}

\caption{Model of the irradiated disc with no Comptonized tail
(i.e. $L_{\rm c}$ = 0), as a function of irradiation fraction,
$f_{\rm out}$ (left panel) and outer radius, $r_{\rm out}$ (right
panel). The models (black solid lines) are calculated for
$kT_{\rm in}$ = 0.6 keV. The left panel shows the effect of
changing the fraction of the bolometric disc emission which is
intercepted by the outer disc, from  0 (the lowest solid black
line) to 0.02, for an assumed outer disc radius of
$10^{4.5}R_{\rm in}$. The right panel shows the effect of
changing the outer disc radius from (right to left) $10^{0.5}$ to
$10^6R_{\rm in}$ for a fixed illumination fraction of 0.01. The
dotted grey line in both panels show a 0.6~keV {\sc diskbb} model
(equivalent to $R_{\rm out}=\infty$) for comparison.}
\label{fig:model_diskir}
\end{figure}

GDP08 showed how irradiation of the inner disc by the Compton tail
could significantly change the temperature structure of the inner disc
in the low/hard state. Neglecting this effect leads to an
underestimate of the true inner disc radius compared to models which
assume that gravitational energy release alone is responsible for the
disc emission (R07). Here we concentrate on how this total emission
(irradiated inner disc and Compton tail) can change the temperature
structure of the outer disc by irradiation to produce the observed
optical/UV emission.

We assume that the disc extends from $R_{\rm in}$ to $R_{\rm
out}$, and model the emission from gravitational energy release
through viscosity using the simple {\sc diskbb} approach
(stressed inner boundary condition) so that
\[
Q_{\rm visc}=\sigma T_{\rm disc}^4(R) =
{GM\dot{M} \over 8\pi R^3} = {L_{\rm d} \over 4\pi R_{\rm in}^2}
r^{-3} = \sigma T_{\rm in}^4 r^{-3},
\]
where $r \equiv R/R_{\rm in}$ and
\[
L_{\rm d} = {GM\dot{M} \over 2R_{\rm in}}
\]
is the Newtonian expectation for the thin disc emission from
gravitational energy release and
\[
T_{\rm in}^4 = {L_{\rm d} \over 4\pi\sigma R_{\rm in}^2}
\]
is the temperature at the inner disc radius as parameterized in
{\sc diskbb} model.

GDP08 considered the effect of irradiation of the disc by the
Compton tail, with luminosity $L_{\rm c}$. This irradiation will
be most concentrated in the overlap region between the disc and
hot flow, so given the uncertainties in geometry, GDP08 simply
assumed that the irradiating flux was constant between $R_{\rm
in}$ and $R_{\rm irr}$ and zero elsewhere. Then the reprocessed
luminosity $L_{\rm rep}=f_{\rm in}L_{\rm c}$, where $f_{\rm in}$
can be estimated via the albedo of the inner disc, $a_{\rm in}$
and the solid angle subtended by the overlap region to the X-ray
source, $\Omega_{\rm in}/2\pi$ as $f_{\rm in}= (1-a_{\rm
in})\Omega_{\rm in}/4\pi \sim 0.1$, for typical hard state values
of $a_{\rm in}=0.3$ and $\Omega_{\rm in}/2\pi=0.3$ (GDP08).  This
gives a flux at each irradiated part of the disc as
\[
Q_{\rm rep} = f_{\rm in} {L_{\rm c} \over 2\pi (R_{\rm irr}^2-R_{\rm in}^2)}
= 2 \sigma T_{\rm in}^4 f_{\rm in} {L_{\rm c} \over L_{\rm d}} {1 \over r_{\rm irr}^2 - 1},
\]
where $r_{\rm irr} \equiv R_{\rm irr} / R_{\rm in}$.

Thus the total luminosity from the inner regions of the accretion
disc is
\[
L_{\rm bol}=L_{\rm d} + L_{\rm rep} + L_{\rm c} = L_{\rm d}\left[1 + {L_{\rm c} \over L_{\rm d}}(1 + f_{\rm
in})\right].
\]
This can in turn irradiate the outer disc, either directly if the
disc shape is convex or warped, or by scattering if there is
material above the disc (such as the wind which is produced self
consistently by X-ray irradiation e.g. Begelman, McKee \& Shields
1983).  We assume the illuminating flux is $f_{\rm il} L_{\rm
bol}/(4 \pi R^2)$, where $f_{\rm il}$ is a geometry-dependent
illumination fraction (see also Dubus et al. 1999; Esin et al.
2000; Hynes et al. 2002). The $R^{-2}$ gives a slower decline
with radius than the $R^{-3}$ dependence of the flux from
intrinsic gravitational energy release. Thus at large radii
irradiation must dominate over the intrinsic gravitational energy
release (e.g. van Paradijs \& McClintock 1994). Of this
illuminating flux, a fraction $a_{\rm out}$ is reflected ($a_{\rm
out}$ is the albedo of the outer disc), and a fraction $\eta_{\rm
th}$ of the remaining absorbed flux is thermalized. We define the
reprocessed fraction
\begin{equation}
f_{\rm out}=f_{\rm il}(1-a_{\rm out}) \eta_{\rm th}
\label{eq:fout}
\end{equation}
as the fraction of the bolometric flux which is thermalized in
the disc. Finally, the total flux in the disc can be described by
the following formula:
\begin{eqnarray*}
Q_{\rm tot} & = & \sigma T^4_{\rm in} \left\{  r^{-3} + f_{\rm out} r^{-2} \left[ 1 + {L_{\rm c} \over L_{\rm d}}(1 + f_{\rm in}) \right] \right. \\
  & + & \left. \delta 2 f_{\rm in} {L_{\rm c} \over L_{\rm d}} {1 \over r_{\rm irr}^2-1} \right\}.
\end{eqnarray*}
Here $\delta$ = 1 for $r \leq r_{\rm irr}$ and 0 otherwise.

This full model\footnote{this model, {\tt diskir}, is publicly
available on the {\sc xspec} web page
http://heasarc.gsfc.nasa.gov/docs/xanadu/xspec/newmodels.html} has 8
free parameters. There are 4 defining the shape of the intrinsic
spectrum (intrinsic disc temperature, Comptonization slope and
electron temperature, and ratio of power in the disc and
Comptonization components), and 2 defining the inner disc reprocessing
(irradiation radius and fraction of flux intercepted). These are all
constrained directly by the X-ray data, while the remaining 2
defining the outer disc reprocessing (outer radius of the disc and
fraction of flux intercepted) are determined from the observed
optical/UV emission.

We first set $L_{\rm c}/L_{\rm d}=0$ to investigate the behaviour
of pure disc self illumination. Fig. \ref{fig:model_diskir} shows
the effect of changing $f_{\rm out}$. The lowest solid curve
shows the pure disc spectrum, while the dotted curve shows a
comparison with {\sc diskbb} at the same temperature and inner
disc radius. These are identical except at the lowest energies,
as the bandpass is so wide that the outer disc radius (fixed at
$\log r_{\rm out}=4.5$ in this plot) becomes important.
Irradiation creates a characteristic `shoulder' with flux
$\propto f_{\rm out} L_{\rm bol} \nu^{-1}$ in the optical/UV.
This shoulder extends between energies $3kT(r_{\rm trans})$ and
$3kT(r_{\rm out})$, where $r_{\rm trans}$ is the radius at which
there is the transition between irradiation and gravitational
energy release Below the shoulder, the flux is simply determined
by the Rayleigh-Jeans tail of the irradiated outer disc so has
flux $\propto \nu^2 T(r_{\rm out}) \propto (f_{\rm out} L_{\rm
bol} / r^2_{\rm out})^{1/4}$. Fig. \ref{fig:model_diskir}b shows
the effect of changing $r_{\rm out}$. Irradiation has no effect
on the disc emission if the disc is smaller than the radius at
which irradiation dominates, whereas for a large disc the
`shoulder' can extend down into the IR.

\begin{figure}
\begin{center}
\leavevmode  \epsfxsize=5cm \epsfbox{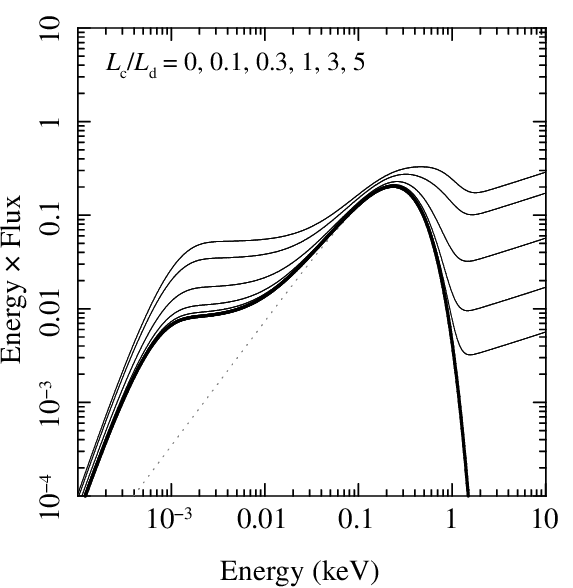}
\end{center}

\caption{Model of the irradiated disc as a function of
Comptonization-to-disc ratio, $L_{\rm c}/L_{\rm d}$. The model
was calculated for $kT_{\rm in}$ = 0.1 keV, $f_{\rm out}$ =
10$^{-2}$, $r_{\rm ir} = 1.1$, lg $r_{\rm out}$ = 4.5,
$\Omega/2\pi$ = 0.3, $\Gamma$ = 1.7 and $kT_e$ = 100 keV. Solid
lines show the models as a function of increasing parameter, from
bottom to top (thicker curve corresponding to $L_{\rm c}/L_{\rm
d} = 0$). The dotted grey line shows {\sc diskbb} model, i.e. the
disc with no irradiation and extending to infinity. Note, that in
a typical soft state ($L_{\rm c}/L_{\rm d} \ll 1$), the UV
shoulder very weakly depends on the strength of the Comptonized
tail, $L_{\rm c}/L_{\rm d}$, but it strongly depends on $f_{\rm
out}$ (see Fig. \ref{fig:model_diskir}).} \label{fig:model_thir}
\end{figure}

We then extend the model to show the effect of including the
Comptonization component on the outer disc illumination,
including also its effect on the inner disc with $r_{\rm
irr}=1.1$ and $f_{\rm in}=0.1$. The total bolometric luminosity
(before reprocessing) is then $L_{\rm d}+L_{\rm c}$, and we
assume that both disc and coronal flux are reprocessed in the
same way, i.e. that a fraction $f_{\rm out}$ of the bolometric
luminosity is thermalized in the outer disc. Fig.
\ref{fig:model_thir} shows a sequence of {\sc diskir} models for
various values of $L_{\rm c}/L_{\rm d}$. When $L_{\rm c}/L_{\rm
d}=0$ (thick blue solid line), the UV shoulder is produced by the
inner disc, $L_{\rm d}$, irradiating the outer disc as in Fig.
\ref{fig:model_diskir}. The height of this shoulder does not
strongly depend on $L_{\rm c}/L_{\rm d}$ in the soft state, as
this has $L_{\rm c}/L_{\rm d}<1$ (by definition of soft state
spectra). Hence the total bolometric luminosity is dominated by
the constant $L_d$, so the reprocessed flux is likewise constant
for a given $f_{\rm out}$ and $r_{\rm out}$. However, for $L_{\rm
c}/L_{\rm d}>1$ then the bolometric luminosity increases
substantially above $L_{\rm d}$, so the reprocessed optical/UV
emission likewise increases with the increase in illuminating
flux.

\section{Results} \label{sec:results}

\begin{table*}
\begin{tabular}{cccccccc}
\hline
$N_H$ & $E_{B-V}/N_H$ & $kT_{\rm in}$ & $f_{\rm out}$ & lg $r_{\rm out}$ & $N_{\rm d}$ & $L_{\rm c}/L_{\rm d}$ & $\chi^2_\nu$/d.o.f.\\
($\times$10$^{21}$ cm$^{-2}$) & $\times$(10$^{-22}$ cm$^{2}$ mag) & (keV) & ($\times10^{-3}$) & & ($\times10^{3}$)\\
\hline
1.16$\pm$0.04 & (1.3)       & 0.890$\pm$0.016  & 0.69$\pm$0.05          & 4.50$\pm$0.05 & 2.62$\pm$0.16 & 0.11$\pm$0.03 & 601.3/542\\
1.16$\pm$0.04 & (1.5)       & 0.890$\pm$0.016  & 0.89$\pm$0.07          & 4.46$\pm$0.05 & 2.62$\pm$0.16 & 0.11$\pm$0.03 & 605.4/542\\
1.16$\pm$0.04 & (1.7)       & 0.890$\pm$0.016  & 1.2$\pm$0.1            & 4.44$\pm$0.04 & 2.62$\pm$0.16 & 0.11$\pm$0.03 & 610.3/542\\
1.16$\pm$0.04 & 0.9$\pm$0.3 & 0.890$\pm$0.016  & 0.36$_{-0.11}^{+0.20}$ & 4.6$\pm$0.1   & 2.65$\pm$0.16 & 0.11$\pm$0.03 & 596.1/541\\
\hline
\end{tabular}
\caption{Fit results of the irradiated disc model to observation
1 (UVOT and XRT spectra) showing the effect of reddening. The
first three fits had $E_{B-V}$ linked to $N_H$, while in the last
fit $E_{B-V}$ was free. The best-fitting value of the colour
excess in the last fit was $E_{B-V} = (0.10\pm0.03)$ mag. $N_H$
is the Hydrogen absorbing column, $E_{B-V}$ reddening, $T_{\rm
in}$ the intrinsic disc temperature at the inner radius, $f_{\rm
out}$ the irradiation fraction, $r_{\rm out}$ the outer disc
radius in units of inner radius, $N_{\rm d}$ is the disc
component normalization (the same as in {\sc diskbb}) and $L_{\rm
c}/L_{\rm d}$ is the ratio of luminosity in the Comptonized
component to that in the non-irradiated disc (i.e. intrinsic disc
emission).} \label{tab:redden}
\end{table*}

We now use the irradiated disc model described in Section
\ref{sec:model} to fit the data. We follow GDP08 and fix the shape of
the Compton tail using the spectral index derived from the RXTE data
and fix its electron temperature to 50~keV (thus the {\sc dthir} model
has only 6 free parameters). We start with observation 1, as it has
the best signal-to-noise and also a full set of filters for the
optical/UV coverage (V, B, U, UVW1, UVM2 and UVW2 i.e. spanning
$\sim$2--8 eV). Absorption by gas and dust in the interstellar medium
affects the X-ray and UV/optical bandpass, respectively (described by
the {\sc xspec} model {\sc wabs} and {\sc redden}). These two should
be linked by a common dust--to--gas ratio, so we first explore the
effect of this on the models.

Gorenstein (1975) found the relation between optical extinction,
$A_{V}$, and Hydrogen column, $N_H =
(2.2\pm0.3)\times10^{21}A_{V}$ cm$^{-2}$ mag$^{-1}$ (see also
Zombeck 1990). Taking the usual relation between extinction and
colour excess (reddening), $A_{V} = 3.0E_{B-V}$, this gives
\[
E_{B-V}/N_H = (1.5\pm0.2)\times10^{-22} {\rm cm}^2 {\rm mag}.
\]
Other values quoted in literature are usually consistent with the
above, e.g. Spitzer (1978) suggests $E_{B-V}/N_H =
1.7\times10^{-22}$ cm$^2$ mag. We fit the first observation with
three fixed values of $E_{B-V}/N_H$ = 1.3, 1.5 and
1.7$\times$10$^{-22}$ cm$^{2}$ mag. In the fourth fit we allow
for $E_{B-V}$ to be a free parameter (independent of $N_H$). The
results are shown in Table \ref{tab:redden}.

The model curves in Fig. \ref{fig:model_diskir} show that the
main parameter controlling the amount of optical/UV flux is
$f_{\rm out}$, while $r_{\rm out}$ controls the optical/UV slope.
Reddening has a fixed shape, so the amount of reddening mainly
affects the the total amount of flux, so should have most impact
on $f_{\rm out}$ This is supported by the results in Table
\ref{tab:redden}, which show that the particular choice of colour
excess has a large effect on the derived outer disc irradiation
fraction, $f_{\rm out}$, and a much smaller effect on $r_{\rm
out}$. It has no effect on any of the inner disc parameters or on
$N_H$ since these are set by the X-ray spectrum.

The clear tradeoff between the irradiation fraction and colour
excess means that the absolute value of $f_{\rm out}$ cannot be
constrained to better than within a factor $\sim$2. However,
relative values can be constrained rather well assuming that
there is no change in the absorption during the outburst. Hence
we fix $E_{B-V}/N_H = 1.5\times10^{-22}$ cm$^2$ mag so that we
can inspect trends in $f_{\rm out}$ throughout the outburst. We
stress that in a typical soft state the UV
shoulder is very weakly affected by the strength of the
Comptonized tail (Fig. \ref{fig:model_thir}), so the measured
values of $f_{\rm out}$ do not depend on a particular value of
$L_{\rm c}/L_{\rm d}$, as long as it remains small (i.e. $\la
0.2$).

We repeated this analysis for observations 2 and 3 as these are the
only other observations with more than 2 filters present.  Thus these
are the only other spectra which can give constraints on the slope of
the optical/UV spectrum and hence on $r_{\rm out}$. These are both
consistent with the $\log r_{\rm out}=4.5$ as found from observation
1, so we fix the outer disc radius at this value in all the datasets.

The outer disc radius of $r_{\rm out} \equiv R_{\rm out}/R_{\rm in}
\approx 10^{4.5}$ translates to a size scale of $3\times10^{11}$~cm
assuming a black hole mass of $M$ = 10 M$_\odot$ and an inner disc
radius of $6GM/c^2$. The disc size is set by tidal forces, which
truncate the disc at roughly half the binary separation. Thus we can
estimate that the orbital period should be about 20 hours, around the
median period for the black hole binary systems (e.g. Remillard \&
McClintock 2006), and requiring a somewhat evolved companion star in
order to fill its Roche lobe.

\begin{figure*}
\begin{center}
\leavevmode  \epsfxsize=13cm \epsfbox{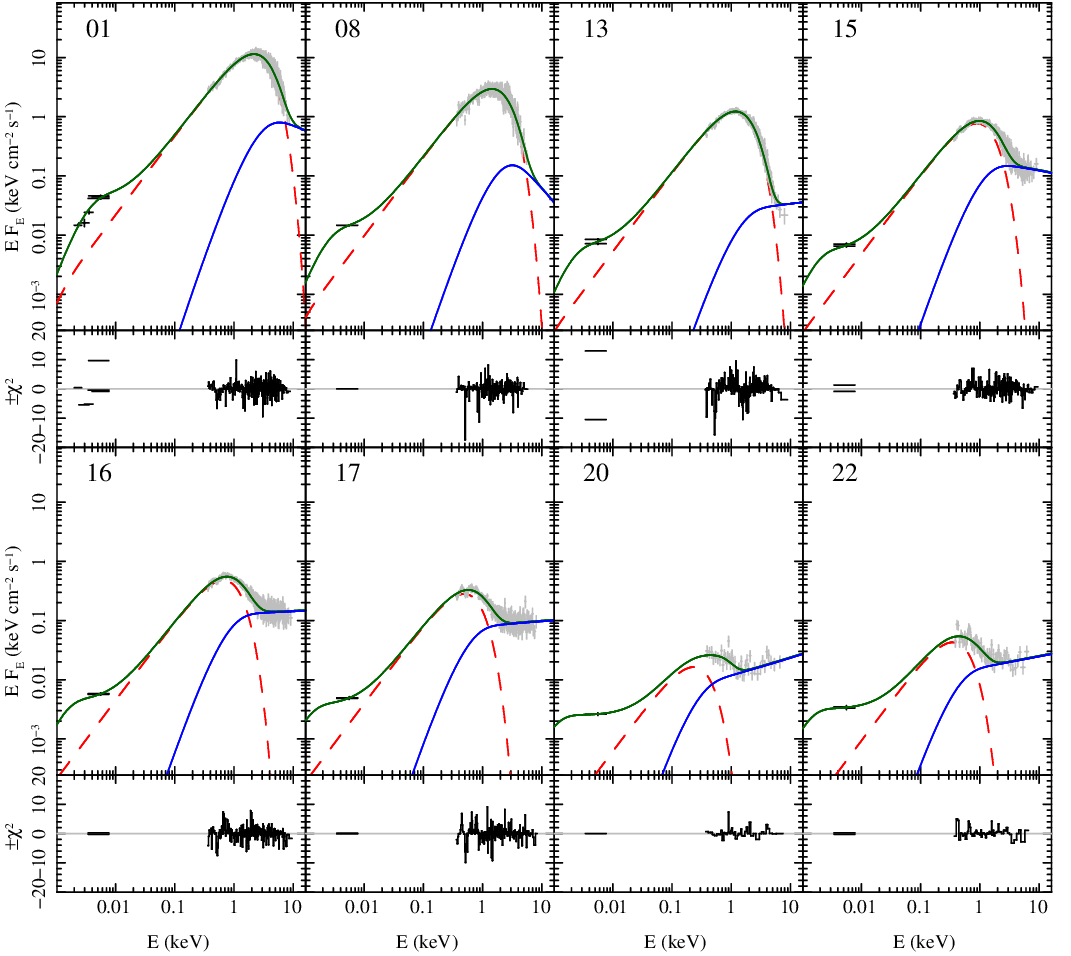}
\end{center}
\caption{Representative spectra from UVOT and XRT with the best
fitting models of the irradiated disc. Upper panels show the
unfolded and unabsorbed spectra and models. Lower panels show the
residuals, $\pm\chi^2 \equiv$ sign(data-model)$\chi^2$. In
spectral panels UVOT data are shown in black, XRT data are in
gray. The dashed red curve shows the intrinsic disc emission due
to accretion (i.e. disc with no irradiation). The blue curve
shows Comptonization, while the dark green curve the total
spectrum, including Comptonization and irradiated disc.
Observations 1, 8 and 13 were in the soft state, observations 15,
16 and 17 show the transition into the hard state, which is
represented by the last two data sets 20 and 22.}
\label{fig:spectra}
\end{figure*}

\begin{table*}
\begin{tabular}{ccccccccccc}
\hline
Obs & $N_H$ & $kT_{\rm in}$ & $f_{\rm out}$ & $N_{\rm d}$ & $\Gamma$ & $L_{\rm c}/L_{\rm d}$ & $\chi^2_\nu$/d.o.f.\\
& ($\times$10$^{21}$ cm$^{-2}$) & (keV) & ($\times10^{-3}$) & ($\times10^{3}$) \\
\hline
 1 & $1.15\pm0.03$          & $0.894\pm0.015$           & $0.86_{-0.04}^{+0.05}$  & $2.6_{-0.1}^{+0.2}$   & (2.34) &  $0.10\pm0.03$        & 607.1/543 \\
 2 & $1.13_{-0.03}^{+0.02}$ & $0.808_{-0.008}^{+0.009}$ & $0.99\pm0.03$           & $2.3\pm0.1$           & (2.28) &  $0.05\pm0.01$        & 677.5/531 \\
 3 & $1.18\pm0.03$          & $0.78\pm0.01$             & $1.06\pm0.04$           & $2.6\pm0.1$           & (2.34) &  $0.05\pm0.02$        & 558.4/499 \\
 4 & $1.12_{-0.03}^{+0.04}$ & $0.70\pm0.01$             & $1.01\pm0.04$           & $3.0\pm0.2$           & (2.31) &  $0.17\pm0.02$        & 520.0/488 \\
 5 & $1.23\pm0.04$          & $0.69\pm0.01$             & $1.08\pm0.06$           & $3.3\pm0.2$           & (2.33) &  $0.19\pm0.02$        & 476.6/452 \\
 8 & $1.07_{-0.08}^{+0.09}$ & $0.60_{-0.02}^{+0.03}$    & $1.00_{-0.07}^{+0.08}$  & $3.3_{-0.5}^{+0.6}$   & (3.00) &  $0.05\pm0.04$        & 352.5/289 \\
 9 & $1.08\pm0.05$          & $0.61\pm0.01$             & $0.99\pm0.06$           & $2.9\pm0.2$           & (3.00) &  $0.01\pm0.01$        & 332.0/340 \\
10 & $1.03\pm0.08$          & $0.53_{-0.01}^{+0.02}$    & $1.08_{-0.08}^{+0.09}$  & $3.2\pm0.4$           & (1.99) &  $0.06\pm0.03$        & 297.8/260 \\
11 & $1.00\pm0.04$          & $0.524\pm0.008$           & $0.95_{-0.05}^{+0.06}$  & $3.7_{-0.2}^{+0.3}$   & (2.74) &  $0.04\pm0.01$        & 363.3/336  \\
12 & $0.95\pm0.03$          & $0.512\pm0.005$           & $0.98\pm0.05$           & $3.3_{-0.1}^{+0.2}$   & (2.23) &  $0.04\pm0.01$        & 422.6/361 \\
13 & $0.96\pm0.03$          & $0.498\pm0.005$           & $1.21_{-0.05}^{+0.06}$  & $2.9\pm0.1$           & (1.88) &  $0.06\pm0.01$        & 530.1/360 \\
15 & $1.12\pm0.05$          & $0.381\pm0.007$           & $1.24\pm0.08$           & $5.5\pm0.5$           & (2.15) &  $0.36\pm0.02$        & 412.6/367 \\
16 & $1.03\pm0.06$          & $0.281\pm0.006$           & $1.27_{-0.07}^{+0.08}$  & $11.5_{-1.3}^{+1.5}$  & (1.95) &  $0.73\pm0.03$        & 408.3/396 \\
17 & $1.0\pm0.1$            & $0.208\pm0.007$           & $1.7\pm0.1$             & $23_{-4}^{+5}$        & (1.92) &  $0.89\pm0.06$        & 366.1/307 \\
20 & $(1.08)$               & $0.10\pm0.02$             & $6.5_{-0.9}^{+1.0}$     & $26_{-13}^{+26}$      & (1.72) &  $4.4_{-1.2}^{+2.8}$  &  39.9/40  \\
22 & $(1.08)$               & $0.15\pm0.02$             & $6.7\pm0.8$             & $13_{-5}^{+8}$        & (1.82) &  $1.6_{-0.2}^{+0.3}$  &  66.1/65  \\
\hline
\end{tabular}
\caption{Fit results of the irradiated disc model, {\sc
redden*wabs*(diskir)}, using all available UVOT/XRT data.
Parameters are as in Table \ref{tab:redden}, with fixed
$E_{B-V}/N_H = 1.5\times10^{-22}$ cm$^{2}$ mag and $\lg r_{\rm
out} = 4.5$. $\Gamma$ is the photon spectral index of the
Comptonized component, derived from the PCA data (see GDP08).}
\label{tab:thir}
\end{table*}

\begin{figure}
\begin{center}
\leavevmode  \epsfxsize=8.5cm \epsfbox{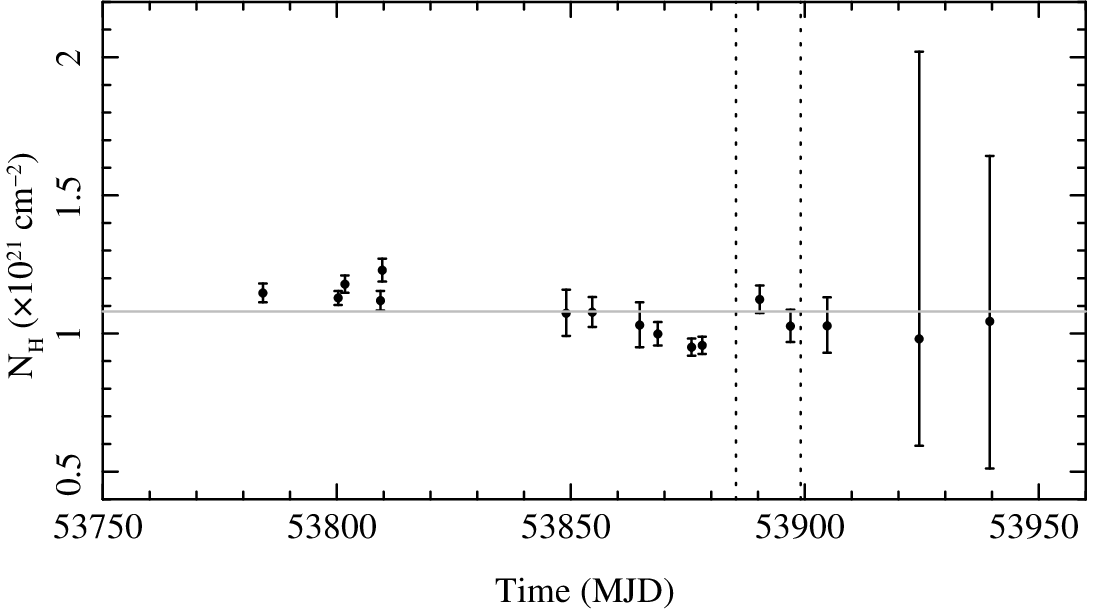}
\end{center}
\caption{The best-fitting absorbing Hydrogen column from the
irradiated disc model. The gray horizontal line shows the mean
value of $N_H = 1.08\times10^{21}$ cm$^{-2}$. Vertical dotted
lines separate the X-ray spectral states: HS is on the left, IM
in the middle and LH on the right.} \label{fig:nh}
\end{figure}

Now we fit the irradiated disc model to all the data sets, with
all free model parameters detailed in Table \ref{tab:thir}. Fig.
\ref{fig:spectra} shows the best-fitting model, data and
residuals of representative spectra. Fig. \ref{fig:nh} shows that
the column density $N_H$ is very well constrained by all the
soft- and intermediate-state data, and is approximately
consistent with a constant value of $N_H =
(1.08\pm0.05)\times10^{21}$ cm$^{-2}$. However, the hard state
observations have much poorer signal to noise, so we fix the
column at this value in the last two data sets in order to better
constrain the remaining parameters.

\begin{figure}
\begin{center}
\leavevmode  \epsfxsize=8.5cm \epsfbox{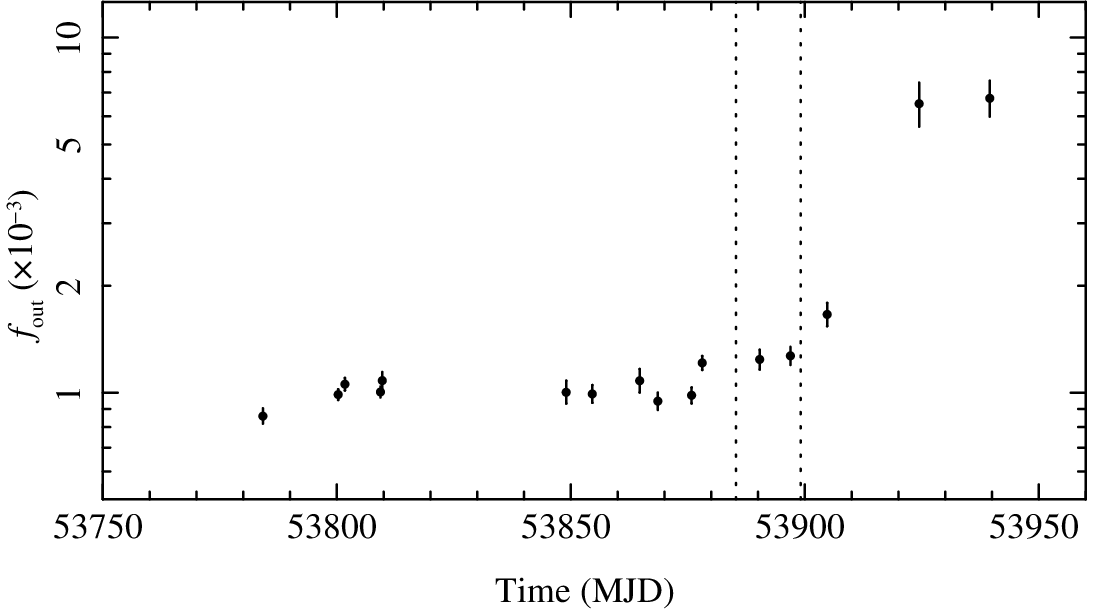}
\end{center}
\caption{Irradiation fraction, $f_{\rm out}$, during the
outburst. Dotted lines separate the X-ray spectral states: HS is
on the left, IM in the middle and LH on the right.} \label{fig:c}
\end{figure}

\begin{figure}
\begin{center}
\leavevmode  \epsfxsize=8.5cm \epsfbox{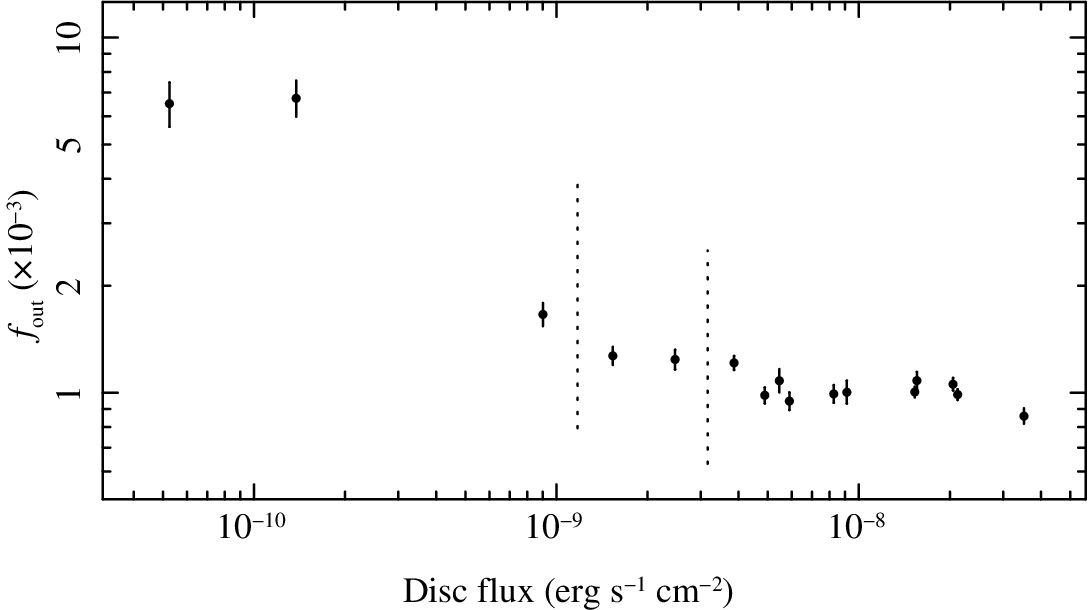}
\end{center}
\caption{Irradiation fraction, $f_{\rm out}$, as a function of
disc flux. Dotted lines separate the X-ray spectral states: HS is
on the right, IM in the middle and LH on the left.}
\label{fig:flux_c}
\end{figure}

\begin{figure}
\begin{center}
\leavevmode  \epsfxsize=6cm \epsfbox{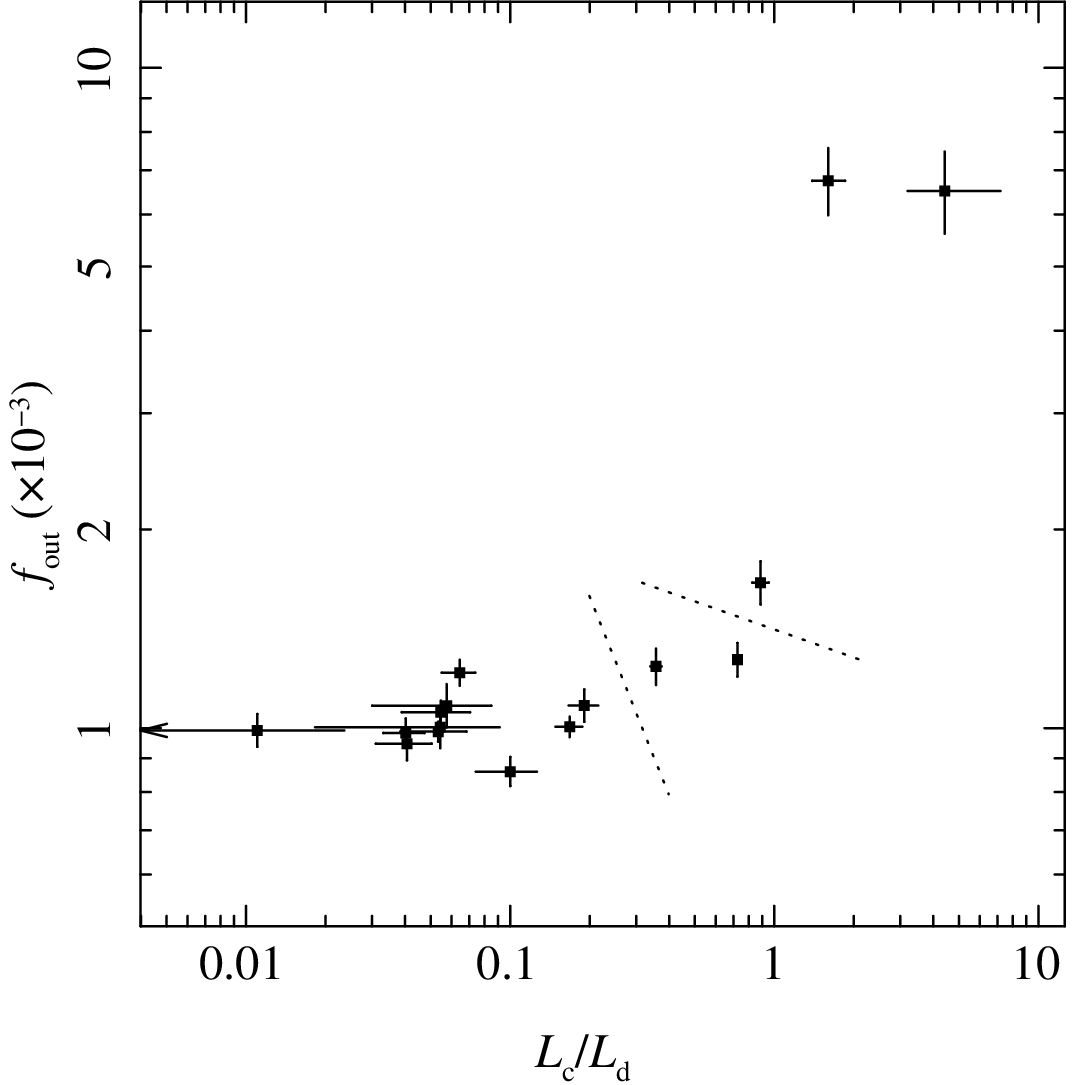}
\end{center}
\caption{Irradiation fraction, $f_{\rm out}$, as a function of
the Comptonization-to-disc ratio, $L_{\rm c}/L_{\rm d}$ from
irradiated disc fitted to UVOT/XRT data.} \label{fig:lcld_fout}
\end{figure}

The irradiation fraction, $f_{\rm out}$, is shown as a function
of time (Fig. \ref{fig:c}), disc flux (Fig. \ref{fig:flux_c}),
and Comptonization--to--disc ratio (Fig. \ref{fig:lcld_fout}). It
remains remarkably constant at $f_{\rm out} \sim 10^{-3}$, during
the soft state and transition, but then jumps by a factor $\sim$6
in the last two hard state spectra. Similar differences between
the optical-X-ray correlation between hard and soft states are
seen by Esin et al. (2000) and Maitra \& Bailyn (2008). We stress
again that while the absolute value of $f_{\rm out}$ is uncertain
to within a factor 2, depending on the assumed reddening, the
relative values are robust. It is also not dependent on the
assumed Compton spectral shape at high energies. Even for the
hardest spectra, with $\Gamma=1.75$, the bolometric luminosity
only increases by a factor 1.3 by changing the high energy extent
from $\sim$150~keV (as produced by a 50~keV electron temperature)
to $500$~keV (more or less the maximum seen in X--ray binaries).
This is much smaller than the factor 6 increase in flux required
in order to make the irradiated fraction be consistent with that
seen in the soft state. It is clear from Fig. \ref{fig:spectra}
that the ratio of the observed UV/optical emission to that
inferred for the underlying intrinsic (unirradiated) disc jumps
markedly for the last two (hard state) spectra, though it is also
clear that the first hard state spectrum (observation 17) is
consistent with the soft and intermediate states.

\begin{figure}
\begin{center}
\leavevmode  \epsfxsize=6cm \epsfbox{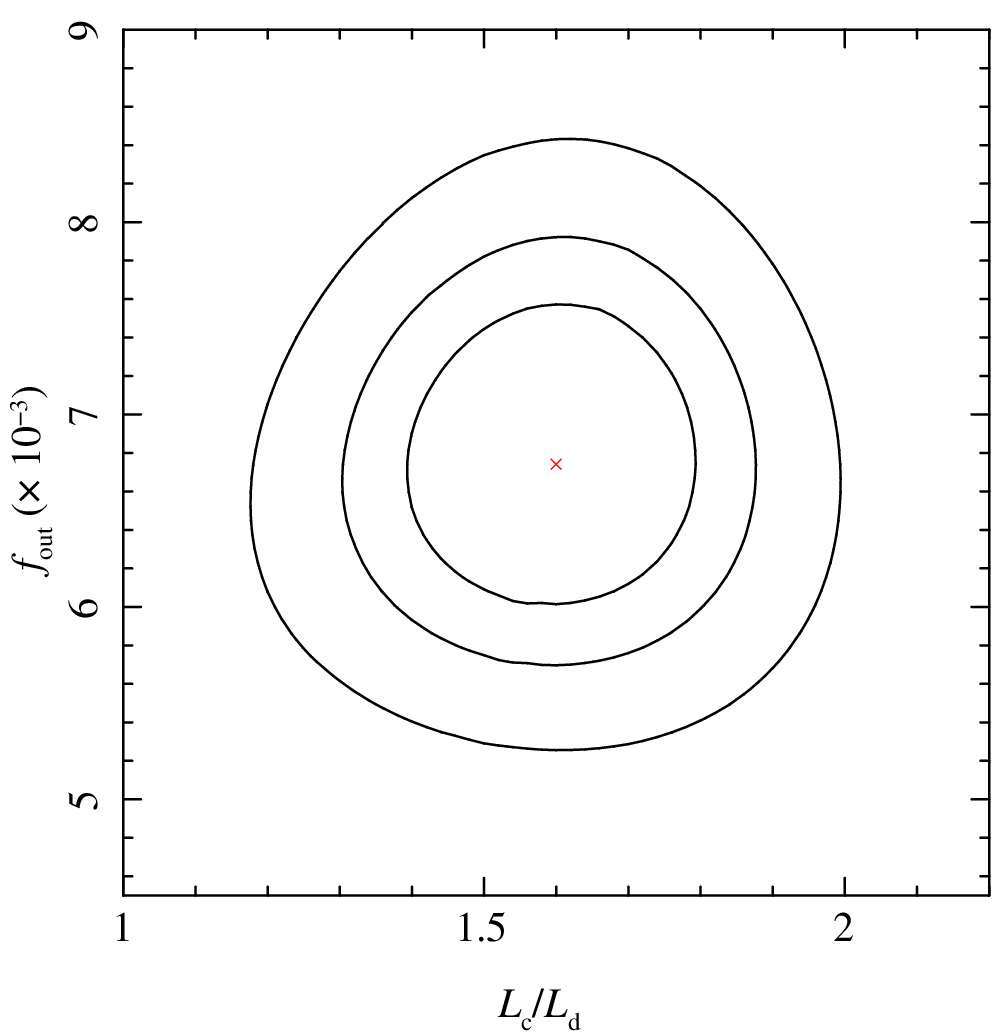}
\end{center}

\caption{Contour plot of $\chi^2$ as a function of  $f_{\rm out}$
and $L_{\rm c}/L_{\rm d}$ for observation 22, using irradiated
disc model (see Fig. \ref{fig:lcld_fout}). The cross shows the
$\chi^2$ minimum and the contours correspond to (innermost first)
68, 90 and 99 per cent confidence level.}\label{fig:cont}
\end{figure}

As discussed in Section \ref{sec:model}, the UV shoulder does not
depend on $L_{\rm c}/L_{\rm d}$ in the soft state. However,
during the transition and in the hard state, where $L_{\rm
c}/L_{\rm d} \ga 1$, the UV shoulder is shaped both by $f_{\rm
out}$ and $L_{\rm c}/L_{\rm d}$ (see Fig.  \ref{fig:model_thir}).
Nonetheless, the X-ray spectral shape constrains $L_{\rm
c}/L_{\rm d}$, and with this then even a single UV data point
independently constrains $f_{\rm out}$. This is the case even for
the hard state, as shown by the lack of correlation between these
parameters in observation 22 (Fig. \ref{fig:cont}), showing that
the hard-state increase in $f_{\rm out}$ is real (Fig.
\ref{fig:lcld_fout} ).

\section{UV/X-ray correlation} \label{sec:correlation}

\begin{figure}
\begin{center}
\leavevmode  \epsfxsize=8cm \epsfbox{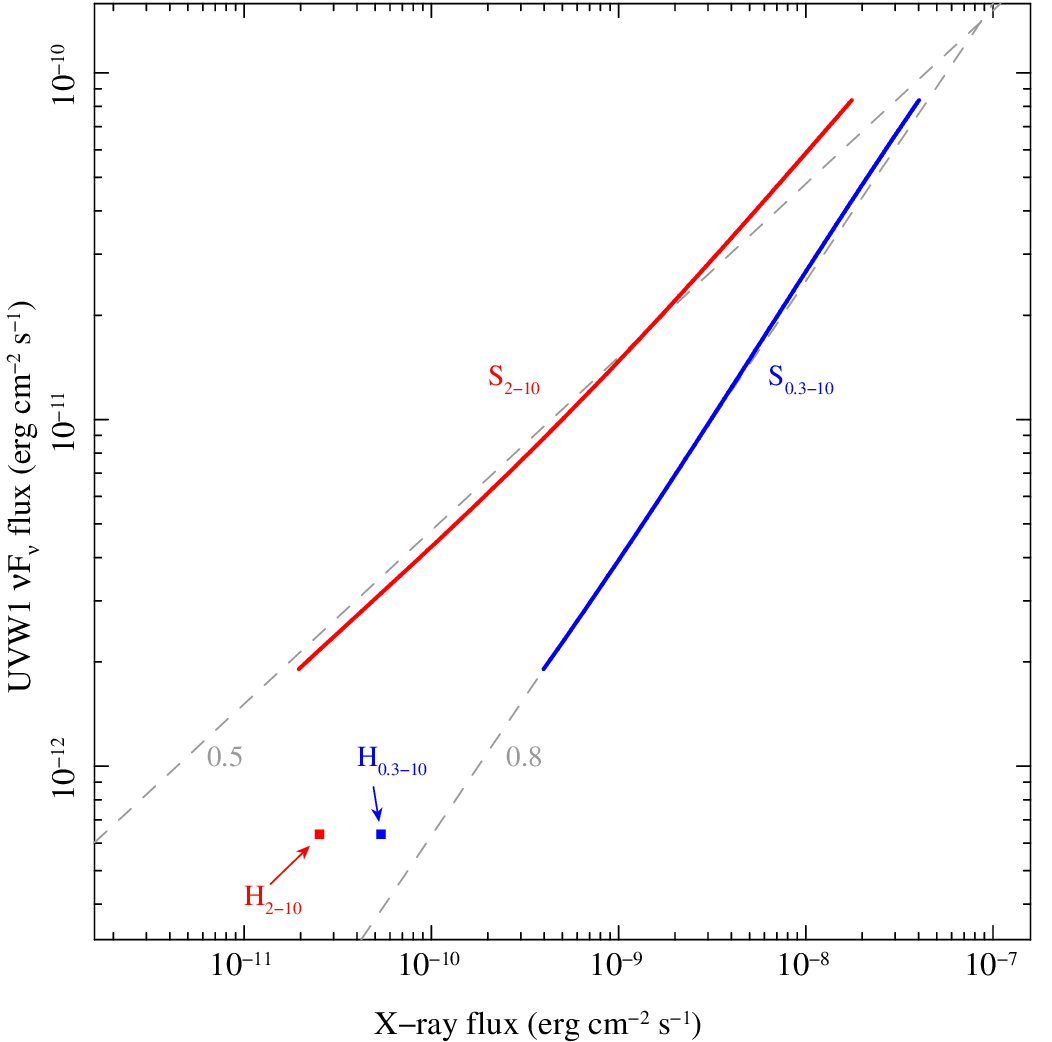}
\end{center}

\caption{X-ray/UV flux correlation simulated using the irradiated
disc model. The UV flux corresponds to the UVW1 filter of {\it
Swift}/UVOT detector, i.e. integrated flux from 3.4 to 7.7 eV.
The X-ray flux is either 2--10 or 0.3--10 keV flux. Gray dashed
lines in the background show power laws with indices 0.5 and 0.8.
The thick curves (marked as S$_{2-10}$ in red and S$_{0.3-10}$ in
blue) represent the soft state correlation. They were modelled
for typical soft-state parameters (see Table \ref{tab:thir}) of
$L_{\rm c}/L_{\rm d}$ = 0.05, $\Gamma$ = 2.2, $N_{\rm d}$ = 3000,
$f_{\rm out}$ = $10^{-3}$ and $kT_{\rm in}$ varying from 0.3 to
0.9 keV. The two points (pointed by arrows and marked as
H$_{2-10}$ and H$_{0.3-10}$) show how the hard state would look
like if it had the same irradiation fraction as the soft state,
$f_{\rm out}$ = $10^{-3}$. The other parameters for the hard
state model were $L_{\rm c}/L_{\rm d}$ = 5, $N_{\rm d}$ = 20000
and $kT_{\rm in}$
 = 0.1 keV.} \label{fig:correlation}

\end{figure}

The previous section showed that there is a robust change in the
ratio of UV/optical to X-ray flux in the hard state compared to
that seen in the soft and transition states, indicating a change
in either the geometry or radiative process (see also Esin et al.
2000). This is initially surprising, as R07 showed that the
2-10~keV X-ray luminosity, $L_{2-10}$, was related to the UV
luminosity in the UVW1 filter as $L_{\rm UVW1}\propto
L_{2-10}^{0.5}$ throughout the outburst, with the hard states
following the same correlation as the soft.  This is the
relationship predicted from reprocessing by van Paradijs \&
McClintock (1994), but only in the special case where the filter
samples the turnover between the `shoulder' (flux $\propto L_{\rm
bol}$) and the Rayleigh-Jeans tail from the outer edge of the
disc ($\propto L_{\rm bol}^{0.25}$, see Section \ref{sec:model}).
Yet the derived disc size in these data is fairly large, and the
UVW1 filter used for the correlation is mostly on the irradiation
dominated shoulder rather sampling the outer disc Rayleigh-Jeans
tail (see Fig. \ref{fig:spectra}). Thus the models predict that
reprocessing should give $L_{\rm UVW1}\propto L_{\rm bol}$ rather
than $\propto L_{\rm bol}^{0.5}$.

However, the 2-10~keV flux is not a good tracer of the bolometric flux.
Even in the soft state, the changing disc temperature means that
the 2-10~keV spectrum is initially dominated by the high
temperature disc while the Compton tail becomes more important as
the disc temperature decreases (see Fig. \ref{fig:spectra}). Thus
there is a variable bolometric correction for the 2--10~keV flux
even in the soft state spectra.

We show all these effects explicitly by simulating the spectrum
of an irradiated disc for typical soft state spectral parameters
(inner and outer radii fixed at $N_{\rm d}=3000$ and $r_{\rm
out}=10^{4.5}$, weak Compton tail of $L_{\rm c}/L_{\rm d} = 0.05$
and $\Gamma=2.2$) with fixed $f_{\rm out}=10^{-3}$ (see Table
\ref{tab:thir}).  The thick curve in Fig.  \ref{fig:correlation}
(marked as S$_{2-10}$) shows the effect of changing the disc
temperature in this model from 0.9~keV to 0.3~keV. This is
approximately described by $L_{\rm UVW1}\propto L_{\rm
2-10}^{0.5}$, while the 0.3-10~keV flux, which more closely
follows the bolometric flux, gives $L_{\rm UVW1}\propto L_{\rm
2-10}^{0.8}$ (actually closer to the `jet signature').  The
approximate `reprocessing signature' of van Paradijs \&
McClintock (1994) is coincidentally recovered in the data by
masking the true reprocessing flux relation by the changing
spectral shape in the X-ray bandpass!

The changing shape of the X-ray spectrum is of course accentuated
during the state transition. Fig.~\ref{fig:correlation} shows the
effect of keeping the irradiated fraction constant ($f_{\rm out} =
10^{-3}$) while changing the spectral shape to a typical hard state
spectrum ($kT_{\rm in}$ = 0.1~keV, $N_{\rm d}$ = 20000, $\Gamma$ =
1.75 and $L_{\rm c}/L_{\rm d}$ = 5). If we used the full bolometric
flux for the correlation, this point should lie on the same line as
the soft-state correlation. However, when the X-ray flux is used (in
particular 2--10 keV flux) the hard state point does not lie on
extrapolation of the soft state (the point marked as H$_{2-10}$ in
Fig. \ref{fig:correlation}).  This change in spectral shape between
hard and soft states then predicts that there is a change in the
optical-X-ray flux correlation measured over a fixed bandpass. Such an
effect is clearly seen by Maitra \& Bailyn (2008) in the neutron star
transient Aql X-1.

The observed hard state, unlike this simulation, follows the
soft-state correlation (R07). This is possible only when there is
an increase in irradiated flux to counteract the large decrease
in bolometric correction. This manifests itself in the increase
of measured irradiation fraction in the hard state (see Fig.
\ref{fig:lcld_fout}).

We {\em strongly} caution against using optical/UV--X-ray
flux-flux correlations as a diagnostic of the origin of the
optical flux. The `reprocessing signature' of $L_{\rm opt}\propto
L_{2-10}^{0.5}$ is only very approximate, depending on the size
of the disc and the changing bolometric correction from
$L_{2-10}$ depending on spectral shape (especially if there are
spectral transitions).

\section{Discussion} \label{sec:discussion}

{\it Swift} observations of XTE J1817--330 during its outburst
clearly show that the UV emission is in excess of a simple
extrapolation of the standard multicolour blackbody disc fitting
the X-ray band (R07). This is as expected from irradiation of the
outer disc by the intense X-ray emission produced in the inner
disc.  We develop a self-consistent model of the X--ray spectrum,
where the broadband optical/UV/X-ray spectrum is made up from
four components, the intrinsic dissipation in the disc, the
Compton tail, irradiation of the inner disc by the tail, and
irradiation of the outer disc by all of the above.

The outer disc can be irradiated in two different ways. Firstly,
a wind or corona can form above the disc and scatter the central
X-rays down. As discussed in section \ref{sec:model}, the
fraction $f_{\rm out}$ of the central luminosity that is
reprocessed, thermalized and reemitted as blackbody (defined by
equation \ref{eq:fout}) depends on the fraction which illuminates
the disc, $f_{\rm il}$, the fraction of this which is absorbed,
$(1-a_{\rm out})$, and the fraction of this absorbed flux which
thermalizes, $\eta_{\rm th}$. In case of the scattering
wind/corona the geometry-dependent factor is
\[
f_{\rm il} = {\Omega_{\rm sc} \over 2\pi} \tau_{\rm sc},
\]
where $\Omega_{\rm sc}$ is the solid angle of the wind/corona
seen by the central source and $\tau_{\rm sc}$ is the optical
depth of the wind/corona along the line of sight from the central
source. In case of direct irradiation the illumination fraction
is simply
\[
f_{\rm il} = {\Omega_{\rm out} \over 2\pi},
\]
where $\Omega_{\rm out}$ is the solid angle of the outer disc
seen by the central source.

The most intriguing result reported in this paper is constancy of
$f_{\rm out}$ in the soft state. The disc temperature changes by
a factor 3, the Comptonization-to-disc ratio varies by an order
of magnitude, the bolometric luminosity changes dramatically,
while the reprocessed fraction remains fairly constant at $f_{\rm
out} \sim 10^{-3}$. This is possible only when the albedo,
$a_{\rm out}$, the thermalization fraction, $\eta_{\rm th}$, and
the illumination fraction, $f_{\rm il}$, remain constant
throughout the soft state. The alternative is a fine tuning of
the above parameters to maintain a constant $f_{\rm out}$.

Here we discuss constancy of these three quantities and draw
possible physical scenarios for the irradiation of the outer disc
in XTE J1817--330 in the soft spectral state.

\subsection{Illumination geometry: soft state}

X-ray illumination heats the top skin of the disc up to the
Compton temperature, and this material escapes as a wind if the
thermal velocity at this temperature is faster than the escape
velocity (Begelman et al. 1983). Simulations show that the column
density and solid angle of this wind give $f_{\rm
il}=(\Omega_{\rm sc}/2\pi) \tau_{\rm sc} \sim L/L_{\rm Edd}$
(Woods et al. 1996). However, this wind is produced only at large
radii, above $\sim 2\times 10^{10} T^{-1}_{\rm IC,8}$~cm for a
$10M_\odot$ black hole, where $T_{\rm IC,8}$ is the Compton
temperature in units of $10^8$~K (Begelman et al. 1983). The
Compton temperature is about a half of the disc temperature for
soft state spectra, so even at the outburst peak with the disc at
0.9~keV this gives $T_{\rm IC,8}\sim 0.05$.  That makes the
radius at which the wind can be produced comparable to our
estimate of the outer disc radius of $3\times10^{11}$ cm. Hence,
it seems likely that the disc is too small to effectively drive a
powerful wind.

Instead, an X-ray heated (static) corona develops above the disc.
It can still scatter flux down onto the disc and this scattered
fraction should still strongly depend on the irradiating flux
i.e. be $\propto L/L_{\rm Edd}$.  Fully iterative calculations of
the self consistent vertical structure produced by an X-ray
illuminated disc are complex (e.g. Jimenez-Garate, Raymond \&
Liedahl 2002), but this material is observed in the highly
inclined neutron star systems (the accretion disc corona and
dipping sources). The compilation of dipping sources by
D{\'{\i}}az Trigo et al. (2006) shows a persistent (non-dip)
column density of $\sim$(4--10)$\times 10^{22}$ cm$^{-2}$ of
photoionized material along the line of sight for neutron stars
with $L/L_{\rm Edd}\sim$ 0.02--0.2. The solid angle of this
material is probably $\sim 0.1$ (e.g. Frank, King \& Lasota
1987), hence the illumination fraction should be $(\Omega_{\rm
sc}/2\pi)\tau_{\rm sc}\sim$ (3--7)$\times 10^{-3}$. This number
can be consistent with the observed $f_{\rm out}$, depending on
the actual value of $a_{\rm out}$ and $\eta_{\rm th}$ (see
Section \ref{sec:albedo}), but it should {\em decrease} by an
order of magnitude as the luminosity declines. This contradicts
the observations.

Hence it seems more likely that the disc is directly illuminated,
where the constant illumination fraction, $\Omega_{\rm
out}/2\pi$, is a natural consequence of a constant disc geometry.
Detailed calculations show that the (steady state) disc shape is
convex, so a pure disc cannot self-illuminate (Dubus et al.
1999), but the self-consistent X-ray illuminated disc vertical
structure is more complex, and can lead to a thicker disc
(Jimenez-Garate et al. 2002; Loska, Czerny \& Szczerba 2004).
Alternatively, the disc can be warped, so that the outer edge is
directly illuminated. Such warps are inferred from the long term
periods seen in some X-ray binary light curves (e.g Tananbaum et
al. 1972) and may be produced simply by  irradiating the disc
(Petterson 1977; Pringle 1996).

\subsection{Albedo and thermalization fraction: soft state}
\label{sec:albedo}

The reflection albedo and fraction of non-reflected flux which
can thermalize are both strongly depend on the vertical structure
of the disc. Material at the local blackbody temperature is
effectively neutral in terms of soft X-ray opacity, so soft
X-rays are absorbed rather than reflected.  Thus the decreasing
disc temperature as the soft state declines predicts a decreasing
reflection albedo. We can estimate the albedo for any of our
model spectra by convolving them with the neutral reflection code
{\sc reflect} in {\sc xspec}. The range of soft-state spectra
used in Fig. \ref{fig:correlation} (i.e. decreasing temperature
from 0.9 to 0.3~keV with constant $L_{\rm c}/L_{\rm d}$ = 0.05)
give an albedo which decreases from $\sim$0.02 to 0.008. These
albedos are so small that the change in absorbed flux, which is
$\propto (1-a_{\rm out})$, is negligible.

This does not necessarily imply that the reprocessed flux should
be a constant fraction of the bolometric luminosity.  The
strongly increasing opacity at softer X-ray energies seen in
neutral material also means that the softest X-rays are absorbed
at such small depths in the disc that their luminosity does not
thermalize. Instead of forming a blackbody spectrum, the energy
is emitted predominantly as line and recombination continuum
above the Lyman limit. Thus the reprocessed emission in the
optical/UV is only a very small fraction of the total reprocessed
flux. Conversely, hard X-rays penetrate deeper down into the
disc, and thermalize (e.g. Suleimanov et al. 1999).

The thermalization fraction can be estimated as the fraction of
luminosity emitted above 2~keV, $\eta_{\rm th}\sim L_{\rm
2-\infty}/L_{\rm bol}$. This factor changes from 0.4 to 0.05 as
the soft state temperature changes from 0.9 to 0.3~keV (the
highest and lowest temperature in the soft state models of Fig.
\ref{fig:correlation}).  Thus the combined effect of albedo and
thermalization in neutral material means that the reprocessed
flux should change by a factor $\sim (1-a_{\rm out})\eta_{\rm th}
\approx \eta_{\rm th}$, i.e. by a factor 8. Our observation of a
constant $f_{\rm out}$ then requires that the illumination
fraction, $f_{\rm il}$, increases by this amount to offset the
changing thermalization fraction! This seems very contrived.

Instead, it seems more likely that the X-ray illumination produces a
thin ionized skin over the disc. This increases its albedo to close to
unity in the limit where skin is completely ionized so that only the
energy from the very hardest X-rays is not reflected due to Compton
down-scattering.  This conclusion is strengthened as similarly high
albedo is also inferred from the lack of irradiation of the companion
star. This implies that the disc opening angle is at least $15^\circ$,
so the outer disc solid angle is not negligibly small. The only way to
the explain the observed weak optical emission from reprocessing is if
most of the X-rays are reflected (van Paradijs 1983; van Paradijs \&
McClintock 1994).

Only the highest energy X-rays are absorbed by such a highly
reflective disc, and these form only a small fraction of the
illuminating flux in the soft state which may remain constant as the
soft state declines. All this flux thermalizes so $\eta_{\rm th}\sim
1$ and $f_{\rm out}=f_{\rm il}(1-a_{\rm out})\eta_{\rm th} \sim f_{\rm
il}(1-a_{\rm out}) \ll 1$. Our value of $f_{\rm out}=10^{-3}$ implies
that the disc intercepts a fraction $f_{\rm il} \approx 2\times
10^{-2}$ of the total bolometric flux for $a_{\rm out}=0.95$ (van
Paradijs \& McClintock 1994).

\subsection{Hard State}

The constant reprocessed fraction, $f_{\rm out}$, seen in the
soft state is most easily interpreted if the outer disc is
directly illuminated and has a constant albedo and thermalization
efficiency. The latter two conditions are probably easier to
achieve if the disc has an ionized skin (self-consistently
produced by the illumination: Jiminez-Garate et al. 2002). If so,
the spectral change to the hard state will produce a large change
in albedo as the luminosity of the hard state peaks at energies
around 100~keV where Compton down-scattering means that the
energy must be absorbed by the disc.  The minimum albedo for the
hard state spectra is around $\sim$0.6 even for completely
ionized material, thus we expect an apparent increase in
reprocessed flux by a factor $\sim$6 for a soft state albedo of
$\sim$0.95, as observed!  Thus it seems most likely that there is
no change in reprocessing geometry as the source makes a
transition to the hard state, but that there is a change in
albedo.

\section{Conclusions}

We develop a simple physical model of an X-ray illuminated disc,
where some fraction of the bolometric luminosity (both the disc
and Comptonized tail) can illuminate the outer disc. We assume
that such cool material has a reflection albedo which is around
$\sim$0.3 and that all the non--reflected luminosity can
thermalize to the local blackbody temperature. Fitting this to
the broadband optical/UV/X-ray {\it Swift} spectra of the
outburst of the black hole binary XTE J1817--330 shows that the
optical/UV emission in the soft state is consistent with a
reprocessing of a constant fraction of the bolometric X-ray
luminosity. This argues against a scattering origin for the
illumination, as the optical depth/solid angle of a wind/corona
from the disc should decrease as the source luminosity declines.
Instead it is much more likely that the outer disc is directly
illuminated by the central source, and maintains a constant
opening angle throughout the soft state.  However, we estimate
that for neutral material the reflection albedo should be very
small, and the thermalization fraction should change markedly as
the soft state declines, requiring a fine tuned increase of the
irradiation geometry in order to produce the constant reprocessed
fraction.  Instead it seems more likely that there is an ionized
skin which develops over the disc, giving a much higher
reflection albedo and a constant thermalization fraction. If so,
then the change in albedo which results as the source makes the
transition to a hard state spectrum can explain the apparent jump
in fraction of reprocessed flux at this point with a constant
reprocessing geometry. This would argue against a significant
additional contribution from the jet to this hard state
optical/UV emission.

These results highlight the additional physical insight which is
derived from fitting even simple physical models of an
illuminated disc as opposed to flux-flux correlations. We urge
further development of more sophisticated models which properly
include the albedo and thermalization effects now that such
excellent data are available to constrain the illumination
geometry of the outer disc.

\section*{Acknowledgements}

MG and CD acknowledge support through a Polish MNiSW grant
NN203065933 (2007--2010) and STFC Senior Fellowship,
respectively.


\label{lastpage}

\end{document}